\documentclass[sigconf,anonymous=false,nabib=true]{acmart}
\usepackage{algorithm}
\usepackage{algpseudocode}
\usepackage{amsmath}
\usepackage{bbm}
\usepackage{booktabs}
\usepackage[show]{chato-notes}
\usepackage{epsfig}
\usepackage[bottom]{footmisc}
\usepackage{graphicx}
\usepackage{hyperref}
\usepackage{numprint}
\usepackage{tikz}
\usepackage{type1cm}
\usepackage{url}        
\usepackage{xspace}
\usepackage{siunitx}
\usepackage{balance}     

\graphicspath{ {images/} }

\newcommand{\R}[1]{{\small \href{https://reddit.com/r/#1}{\texttt{r/#1}}}}
\newcommand{\Rsmall}[1]{{\footnotesize \href{https://reddit.com/r/#1}{\texttt{r/#1}}}}
\newcommand{\donald}{\R{The\_Donald}\xspace}
\newcommand{\dataset}{\texttt{reddit-politics-12-16}\xspace}

\hypersetup{
    bookmarks=true,         
    unicode=false,          
    pdftoolbar=true,        
    pdfmenubar=true,        
    pdffitwindow=false,     
    pdfstartview={FitH},    
    pdftitle={My title},    
    pdfauthor={Author},     
    pdfsubject={Subject},   
    pdfcreator={Creator},   
    pdfproducer={Producer}, 
    pdfnewwindow=true,      
    colorlinks=true,       
    linkcolor=black,          
    citecolor=black,        
    filecolor=black,      
    urlcolor=black           
}

\newcommand{\spara}[1]{\smallskip\noindent{\bf #1}}
\newcommand{\mpara}[1]{\medskip\noindent{\bf #1}}

\newenvironment {squishlist}
{\begin{list}{$\bullet$}
  { \setlength{\itemsep}{0pt}
     \setlength{\parsep}{3pt}
     \setlength{\topsep}{3pt}
     \setlength{\partopsep}{0pt}
     \setlength{\leftmargin}{1.5em}
     \setlength{\labelwidth}{1em}
     \setlength{\labelsep}{0.5em} } }
{\end{list}}

\newcommand{\savespace}{\vspace{-4mm}}
\newcommand{\savespaceextra}{\vspace{-1mm}}

\usepackage{draftwatermark}
\SetWatermarkText{Please cite the published version (WebSci '20). \linebreak DOI: 10.1145/3394231.3397894}
\SetWatermarkScale{0.08}
\SetWatermarkAngle{0}
\SetWatermarkVerCenter{10mm}

\copyrightyear{2020} 
\acmYear{2020} 
\setcopyright{acmlicensed}\acmConference[WebSci '20]{12th ACM Conference on Web Science}{July 6--10, 2020}{Southampton, United Kingdom}
\acmBooktitle{12th ACM Conference on Web Science (WebSci '20), July 6--10, 2020, Southampton, United Kingdom}
\acmPrice{15.00}
\acmDOI{10.1145/3394231.3397894}
\acmISBN{978-1-4503-7989-2/20/07}

\begin{document}

\title[Roots of Trumpism]{Roots of Trumpism: Homophily and Social Feedback\\in Donald Trump Support on Reddit}

\author{Joan Massachs}
\affiliation{\institution{Universitat Polit\`ecnica de Catalunya, Spain}}
\email{joan.massachs@est.fib.upc.edu}

\author{Corrado Monti}
\affiliation{\institution{ISI Foundation, Italy}}
\email{corrado.monti@isi.it}

\author{Gianmarco De~Francisci~Morales}
\affiliation{\institution{ISI Foundation, Italy}}
\email{gdfm@acm.org}

\author{Francesco Bonchi}
\affiliation{\institution{ISI Foundation, Italy}}
\affiliation{\institution{Eurecat, Spain}}
\email{francesco.bonchi@isi.it}

\renewcommand{\shortauthors}{J. Massachs, C. Monti, G. De~Francisci~Morales, F. Bonchi}

\begin{abstract}
  We study the emergence of support for Donald Trump in Reddit's political discussion.
  With almost 800k subscribers, ``\donald'' is one of the largest communities on Reddit, and one of the main hubs for Trump supporters.
  It was created in 2015, shortly after Donald Trump began his presidential campaign.
  By using only data from 2012, we predict the likelihood of being a supporter of Donald Trump in 2016, the year of the last US presidential elections.
  To characterize the behavior of Trump supporters, we draw from three different sociological hypotheses: homophily, social influence, and social feedback.
  We operationalize each hypothesis as a set of features for each user, and train 
   classifiers to predict their participation in \donald.

  We find that homophily-based and social feedback-based features are the most predictive signals.
  Conversely, we do not observe a strong impact of social influence mechanisms.
We also perform an introspection of the best-performing model to build a ``persona'' of the typical supporter of Donald Trump on Reddit.
  We find evidence that the most prominent traits include a predominance of masculine interests, a conservative and libertarian political leaning, and links with politically incorrect and conspiratorial content.
\end{abstract}


{\mathchardef\UrlBreakPenalty=10000
\maketitle \sloppy
}
\mathchardef\UrlBreakPenalty=\relpenalty

\section{Introduction}
\label{sec:intro}


%

The emergence and success of Donald Trump during the 2016 US presidential elections caught many pundits by surprise.\footnote{\url{https://www.forbes.com/sites/stevedenning/2016/11/13/the-five-whys-of-the-trump-surprise}}
The reasons behind such an upset have been the subject of intense debate:
they have been traced back to a resurgence of authoritarian populism, to the socio-economic context of US in a globalized world, and even to his raw appeal as an anti-establishment and divisive candidate, just to name a few~\citep{de2019agrarian,ahmadian2017explaining,sherman2018personal,fitzduff2017irrational,mutz2018status}.

While understanding the precise causes of Trump's success might be impossible, the unprecedented data available via the Web and social media gives us an opportunity to at least understand his supporters.
Indeed, the goal of this work is to study the emergence of support for Donald Trump in Reddit's political discussion.
Donald Trump's campaign relied heavily on social media, and Reddit was a fundamental platform for its success~\citep{karpf2017digital}.
Moreover, Reddit allows to study this emergence in a broader perspective, by identifying which factors anticipate Trump support years before.

Reddit is a social news aggregation website; in 2012, it attracted 46 million unique visitors; in 2016, it was the seventh most visited website in United States, with more than 200 million visitors.\footnote{\url{http://web.archive.org/web/20121231152526/http://www.reddit.com/about/}\\\url{http://web.archive.org/web/20161213123205/https://www.alexa.com/topsites/countries/US}}
Its users use pseudonyms, and their posts and comments are publicly available.
Reddit is also commonly used to discuss news and political topics.
These features make it a promising venue for social research.
Moreover, one of the largest online communities of Donald Trump supporters is the Reddit community \donald.

Although this community was born only in 2015, thanks to the availability of historical Reddit data over the years, we can frame our investigation as a prediction task.
Thus, our methodology in this work is the following.
First, we build a \emph{computational focus group}~\cite{lin2013voices} of \num{44924} politically active users on Reddit, who engaged in political discussion both in 2012 and in 2016.
Then, we divide our focus group into two classes: those who participate in \donald in 2016 and those who do not.
Participation in \donald is a valid proxy to study Donald Trump support, as the rules of this subreddit explicitly state that the community is for ``Trump Supporters Only'', and that dissenting users will be removed.
Based on this proxy, we identify \num{7083} (15.8\%) users with significant presence in that community. 

Therefore, we frame our question as a binary prediction task: \emph{given the features of a user in 2012, can we predict whether they will participate in \donald in 2016?}

For our purpose, we define a set of features by drawing from existing sociological theories of opinion formation. In particular, our features capture three social mechanisms: influence, conformity, and homophily.
Each mechanism is the product of a different type of interaction between a user and their environment.
First, we consider direct communications---a user paying attention to a comment.
This interaction might lead to attitude change through persuasion or reactance; in general, we speak of \emph{(direct) influence}.
Determining whether online interactions on social media can cause one to reconsider their views has attracted considerable attention~\cite{diehl2016political} and several concerns~\cite{lou2019information}.
The second type of interaction we consider is \emph{social feedback}.
It might lead to attitude change via \emph{conformity}~\cite{cialdini2004social}, since users might wish to match the perceived norm of their communities.
The opposite can also happen: \emph{anti-conformity}~\cite{willis1963two} can lead users to defy the perceived norms they experience.
We operationalize social feedback as the score received by a user in a particular community.
Finally we consider indirect interactions: common interests, proximity, social groups.
They might explain common attitudes via \emph{homophily}~\cite{mcpherson2001birds}.
We observe indirect interactions as participation in Reddit communities.
These are not necessarily political, and include also hobbies, interests, religions, geographic locations, and even addictions.
Distinguishing influence from homophily is a long-standing problem in social network analysis~\cite{la2010randomization}.

By aggregating these three sets of features, we build a rich data set regarding our focus group of politically active users.
We share this data set, dubbed \dataset, for further investigation on this topic.
In this work, we use it to answer the following research questions:
\begin{squishlist}
\item \emph{Can we predict who will support Donald Trump four years in advance?}
\item \emph{Which kind of interaction is most predictive of participation in \donald?}
\item \emph{What are the main traits of a future Trump supporter on Reddit?}
\end{squishlist}

Our best model achieves an F1-score of $35.3\%$, more than double the random baseline of $15.2\%$, and an area under the ROC curve of $0.70$.
We find evidence that homophily is the better predictor among the considered ones, while conformity also plays a noticeable role.
We do not observe significant evidence of direct influence.
Several interesting traits emerge among those that predict Donald Trump support, which we describe in detail in Section~\ref{sec:results}.
The Trump supporter ``persona'' has conservative and libertarian views, and participates in politically incorrect and conspiratorial communities.
Among their interests, the most important ones are entrepreneurship, guns, and video games.
Among the traits more heavily anti-correlated with Trumpism, we find atheism and environmentalism, as well as interests such as cooking and DIY electronics.


\section{Background and related work}
\label{sec:related}


Reddit, as an interesting and publicly available data source, has attracted plenty of attention in recent works.
A comprehensive survey was compiled in 2017 by \citet{medvedev2017anatomy}.
%
More recently, some works have used Reddit data to study the evolution of specific beliefs and tendencies; as well as the relationship between politics and different Reddit communities.
\citet{kane2018communities} use LDA to characterize the political tendencies of non-political subreddits; however, the presence of arguments makes their results hard to interpret.
\citet{klein2019pathways} characterize Reddit users that joined the \R{conspiracy} subreddit, as a proxy to study conspiratorial world views.
They find that language differs clearly between conspiratorial users and their control group; in particular, they observe differences in usage of words related to crime, government and power, while they do not witness meaningful differences in negative or positive emotions.
They also analyze which subreddits act as ``pathways'' to \R{conspiracy} by building a user-based similarity network between communities.
Subscribers of \R{conspiracy} are over-represented in communities related to pornography, tech culture, and music.
As we show in Section~\ref{sec:results}, we find a significant correlation between \R{conspiracy} and \donald.
\citet{grover2019detecting} analyse behaviour patterns in \R{altright}, finding that they display warning behaviors such as fixation and in-group  identification.

A small number of works explicitly focus on \donald.
\citet{zannettou2018origins} study the propagation of memes across multiple alt-right communities in social networks, including Reddit and \donald.
\citet{flores2018mobilizing} investigate the behaviour of users on \donald, finding that they often adopt ``troll slang'', especially when discussing conspiracy theories.
They also find that the messages attracting most engagement are those explaining in detail some political circumstances and calling users to action. 
In their conclusions, they also note the need for a deeper look at this community, by investigating its roots.


\medskip

Our prediction task can be considered related to stance detection, as we identify the opinion of a pre-determined set of individuals with respect to a specific topic.
Usually, however, stance detection involves determining the stance of a short text, typically where the author explicitly mentions the stance object.
For instance, \citet{mohammad2016semeval} at the SemEval-2016 Task~6 challenge classify the stances of a set of Twitter users on different topics.
Interestingly, one of the topics of the challenge is Donald Trump's presidential candidacy, on which the best classifier achieves an F1-score of $0.56$ (compared to a constant baseline F1-score of $0.29$).
The classification performance metrics of our best model are in line with these results.
Other examples of political stance detection include the work by \citet{lai2018stance} about classifying stances on the Italian 2016 referendum, and the one by \citet{taule2017overview} about stances on Catalonian independence.  

Usually, stance detection methods rely heavily on linguistic features~\citep{mohammad2016semeval, taule2017overview, lai2018stance} to predict explicit views.
However, it is also possible to use homophily to identify significant correlation between political beliefs and other traits.
To quote \citet{dellaposta2015liberals}, \emph{``self-reinforcing dynamics of homophily and influence dramatically amplify even very small elective affinities between lifestyle and ideology''}.
This phenomenon has been studied on Twitter by~\citet{garimella2014co}, by analyzing significant traits of democrat and republican Twitter users.
\citet{magdy} employ a mix of these features to predict Islamophobic views on Twitter before they are expressed.
Network features alone are able to achieve a precision of $79\%$ on this task, thus confirming the importance of homophily in predicting unspoken views.


\section{Data}
\label{sec:data}




We take our data set from Reddit~\cite{baumgartner2020pushshift}.
Reddit is organized in topical communities, called \emph{subreddits}.
Users can \emph{post} in these subreddits, and \emph{comment} on other posts and comments, 
thus creating a tree structure for the overall discussion.
We call a \emph{message} a generic piece of user-generated content, when the distinction between post and comment is not relevant.
In addition, users can also \emph{upvote} a message to show approval, appreciation, or agreement (and their opposites with a \emph{downvote}).
The \emph{score} of a message is the number of positive votes minus the number of negative votes it has received.\footnote{\url{https://www.reddithelp.com/en/categories/reddit-101/reddit-basics/how-posts-or-comments-score-determined}} 

To define our focus group, we first need to define the set of subreddits we wish to consider.
Since we are interested in political discussion, we choose \R{politics}, the largest political subreddit, as our seed.
We then pick the $50$ most similar subreddits to \R{politics} according to cosine similarity over a vector representation of the subreddits based on latent semantic analysis, which captures subreddits whose user base is similar to the seed one.\footnote{\url{https://www.shorttails.io/interactive-map-of-reddit-and-subreddit-similarity-calculator}}

By considering these \emph{political subreddits}, let the set of \emph{active users} be those that have written at least $10$ comments in $2012$ and $10$ comments in $2016$ in any of these subreddits.
This set contains \num{44924} users, and constitutes our computational focus group~\cite{lin2013voices}.
In addition, let the \emph{popular subreddits} be the top \num{1000} subreddits with the most comments.


Let us now focus on the task at hand.
We wish to predict which users will support Trump in 2016, the year Trump was elected president of the United States, by looking only at data from 2012, the year of the previous presidential elections.

\paragraph{Class label.}
We use participation in \donald in 2016 to infer the class label of politically active Reddit users.
It is worth mentioning that in 2012 the subreddit \donald did not exist yet, so we have no notion of Trump supporters in 2012.
However, simply taking all users who commented in \donald is too loose and noisy as an operational definition.
As a first approximation, we define a user to be a Trump supporter if they have at least 4 comments on \donald, and the sum of their scores is at least 4.
This corresponds to \num{7427} users; however, we note that \num{1200} of those users have also posted on
the subreddit devoted to the other presidential candidate, Hillary Clinton (\R{hillaryclinton}).
Therefore, in order to take into account the general political activity of a user, we consider a user as Trump supporter in 2016 if they have at least 4 comments more in \donald than in \R{hillaryclinton}, and the sum of the scores (both positive and negative) on \donald is at least 4 points higher than the one in \R{hillaryclinton}.
This definition allows us to have a data set with limited class imbalance while maximizing the confidence in the label attribution.
With this method, we discard \num{344} users (4.6\% of our first set) that, according to this definition, are not clearly supporting Trump in 2016.
Finally, in our focus group of \num{44924} users, \num{7083} (15.8\%) are labeled as Trump supporters and \num{37841} (84.2\%) are labeled as non Trump supporters.
This labeling is what we adopt in all of our analysis.

\paragraph{Direct influence.}
We say that an active user $u$ interacts with the political subreddit $r$ when $u$ answers a message, in any popular subreddit, made by another user $v$ who has posted in the subreddit $r$ in $2012$.
This notion of direct influence captures the idea that $u$ interacts with $v$, who is a user belonging to the community $r$, and therefore is possibly exposed to the attitudes of that community, irrespective of where the interaction takes place.
We opt for this notion of influence to avoid extreme sparsity from considering user to user interactions.

Furthermore, we consider an interaction \emph{conflictual} when one of the two messages has a score of at least $10$ and the other one has a score of at most $-10$.
This definition captures the notion that the two attitudes expressed in the messages differ, and that the interaction possibly represents a conflict.
For each active user and political subreddit, we compute how many times the user has interacted with the subreddit, and how many of these interactions are conflictual.

\paragraph{Social feedback.}
We consider the scores received by an active user $u$ on a political subreddit $r$ in 2012 as a proxy for the social feedback given by $r$ to $u$.
The positive and negative scores are considered separately, as forms of positive and negative reinforcement, respectively.
We use average scores to normalize the score across different levels of user activity.
The higher the average positive score of a user, the better received their attitude is in the given community.
Conversely, the average negative score shows how much a given community disapproves of the attitude of a given user.

\paragraph{Homophily.}
Users may have similar behavior --support Trump-- because they already have similar characteristics and interests.
We capture this notion by looking at the participation of an active user $u$ to a popular subreddit $r$.
Users with similar interests are likely to belong to the same communities, which is a form of homophily.
We experimented with both numerical (number of comments) and binary versions of these features, and found the results to be similar.
Given that the latter version is simpler to interpret, henceforth we report results for the binary feature.


Therefore, our final data set contains the following features for each user:
\begin{squishlist}
  \item[\textbf{Participation:}]
  \item The feature \emph{r part.} is true when the user participates in subreddit $r$, i.e., they have written a comment on $r$.
  \item[\textbf{Score:}]
  \item The feature \emph{r pos. s.} is the average of the positive scores of the comments by the user in subreddit $r$.
  \item The feature \emph{r neg. s.} is the average of the negative scores of the comments by the user in subreddit $r$.
  \item[\textbf{Interaction:}]
  \item The feature \emph{num. i.} is the total number of direct interactions that the user has had.
  \item The feature \emph{r dist. i.} is the fraction of direct interactions that the user has had with users participating to the subreddit $r$. 
  \item The feature \emph{r pos. i.} is the fraction of \emph{non-conflictual} direct interactions with users participating to the subreddit $r$ among the direct interactions with users participating to $r$.
\end{squishlist}
This data set is the main artifact resulting from our research.
We believe it is of independent value for research in computational social science, and thus make it available to the community.\footnote{\url{https://github.com/JoanMG/reddit-data}}

\smallskip

For both scores and positive interactions, if the user does not have comments in the subreddit $r$, and thus the features would be undefined, the value of the feature is taken as the population average.
This way, the classification algorithm cannot distinguish an average score value from a non-participating user.
In other words, this imputation method removes the participation information from the features, with the aim of disentangling homophily from social feedback and direct influence.

\smallskip

In addition, we extract two other sets of interpretable baseline features grounded in text mining:
\begin{squishlist}
  \item[\textbf{Sentiment:}] The feature \emph{r polarity} is the average polarity of the titles of the posts by the user in a political subreddit $r$.
  We compute the polarity by using TextBlob.\footnote{\url{https://textblob.readthedocs.io}}
  \item[\textbf{Bag of words:}]
  The feature \emph{x bag} is the tf-idf weight of the word $x$ in the titles of the posts in political subreddits the user has authored.
\end{squishlist}

Moreover, we create two derived feature sets: bisected scores and bisected interactions.
These features are based on the score and interaction features, by dividing the subreddits in two sets.
The grouping is defined depending on whether the fraction of Trump-supporting users in 2016 is above or below average for the given subreddit.
Let us indicate these two sets of subreddits with $T$ and $N$, respectively.
Given that this feature grouping uses the label information, we do not use them to investigate their predictive power.
Rather, we leverage them to gain insights on which features are correlated with Trump support.
For the \emph{bisected scores}, rather than having a positive and negative value for each subreddit, we have only four values: average positive and negative scores for each of the two groups of subreddits.
Similarly, for \emph{bisected interactions}, the interactions of a user are summarized in three values: ($i$) the fraction of direct interactions that the user has had with users participating in a subreddit in $T$, 
($ii$) the fraction of \emph{non-conflictual} direct interactions with users participating in a subreddit in $T$, and 
($iii$) the fraction of \emph{non-conflictual} direct interactions with users participating in a subreddit in $N$. 


\section{Methods}
\label{sec:methods}

For each feature set described in the previous section, we train different classification algorithms to predict which users will become Trump supporters in 2016.
In addition, we also test the possible combinations between participation, score, and interaction features.

Before training each classification algorithm, we preprocess the data and perform feature selection to avoid overfitting and to obtain more parsimonious and interpretable models.
In particular, we perform the following preprocessing steps: ($i$) remove sparse features, ($ii$) standardize numerical values, ($iii$) select only significantly correlated features, and ($iv$) remove multicollinearity.

In the first step, we remove features that are defined for fewer than \num{500} users (out of \num{44924} total);
for the participation feature set, we use a stricter rule and remove subreddits with fewer than \num{250} users in our group;
for the bag-of-words feature set, we remove the words that are used by fewer than \num{45} users ($0.1\%$ of our focus group).
In the standardization step, we shift and rescale each numerical feature so that it has zero mean and unit variance.
For feature selection, we remove all features that are not significantly correlated ($p < 0.05$) with the target variable, according to Pearson correlation.
Finally, to remove multicollinearity, we iteratively remove the most significantly collinear features through a greedy approach for backward feature elimination; we measure collinearity by means of variance inflation factor (VIF).

After feature selection, we train the following machine learning algorithms: logistic regression, decision tree, and random forest.
For each one, all the measures reported are obtained through 5-fold cross-validation.
We optimize the hyper-parameter of each classification algorithm by using nested cross-validation, so as not overfit the model selection stage.
We report the average F1 measure and the standard deviation across the 5 folds for the best model (according to the nested cross-validation).

\section{Results}
\label{sec:results}

In this section, we present our experimental results and provide answers to our original research questions.
Firstly, we measure and discuss the prediction accuracy of each feature set, to determine how well we can predict Trump support and which kind of interaction is the most predictive.
Secondly, we analyze the most predictive features, to outline the main traits that distinguish future Trump supporters on Reddit.

\subsection{Prediction accuracy}

\begin{table}[t]
  \small
  \centering
  \caption{For each algorithm and for each feature set, we report the F1-score (\%) and its standard deviation $\sigma$ over the 5-fold cross-validation. The three algorithms used are logistic regression, decision tree, and random forest. For a detailed description of each feature set, see Section~\ref{sec:main-results}.}
  \label{tab:main-results}
  \begin{tabular}{@{}lrrrrrrrr@{}}
    \toprule
    & \multicolumn{2}{c}{\textbf{LR}} && \multicolumn{2}{c}{\textbf{DT}} && \multicolumn{2}{c}{\textbf{RF}} \\
    \cmidrule(lr){2-3} \cmidrule(lr){5-6} \cmidrule(lr){8-9}
                     &           F1  (\%)& $\sigma$ &&           F1  (\%)  &  $\sigma$ &&           F1  (\%)  &  $\sigma$ \\
    \midrule
    Participation    & \textbf{34.8} & 0.7 &&         31.8  &  0.5 &&         33.7  &  0.7 \\
    Score           &         29.5  & 1.2 &&         31.0  &  1.7 && \textbf{33.7} &  1.0 \\
    Interaction     & \textbf{26.7} & 0.7 &&         26.3  &  1.0 &&         25.5  &  0.6 \\
    \addlinespace
    Sentiment        &          7.3  & 0.8 && \textbf{16.4} & 13.4 &&         10.7  & 13.1 \\
    Bag of words     & \textbf{25.9} & 1.0 &&         13.1  & 10.7 &&         23.1  &  0.6 \\
    \midrule	[0.5\lightrulewidth]
    Score (bisected)&         29.0  & 0.8 &&         29.5  &  0.8 && \textbf{29.8} &  0.9 \\
    Int. (bisected)  &         25.4  & 0.6 && \textbf{26.9} &  0.8 &&         24.6  &  1.5 \\
    \addlinespace
    Int. + Part.     & \textbf{34.7} & 1.3 &&         31.5  &  0.7 &&         33.8  &  0.8 \\
    Int. + Score    &         30.4  & 1.0 &&         30.6  &  1.8 && \textbf{33.6} &  0.9 \\
    Part. + Score   & \textbf{35.3} & 0.9 &&         32.3  &  0.7 &&         35.0  &  0.6 \\
    Int. + Part. + Score & \textbf{35.5} & 1.2 &&         32.0  &  0.7 &&         35.2  &  0.8 \\
    \midrule
    \multicolumn{9}{c}{Random baseline F1: $15.2\%$} \\
    \bottomrule    
  \end{tabular}
  \savespace
\end{table}

Our results for each feature set and classifier are summed up in Table~\ref{tab:main-results}.
First, note that logistic regression outperforms the other two algorithms in most cases, although there are some exceptions --score-based, sentiment, and bisected interactions features-- that we discuss in the following paragraphs. 
We now compare the predictive power of each feature set by looking at the F1-score achieved by the best classifier. \label{sec:main-results}

\spara{Homophily.} Participation is the best-performing feature among the basic sets; it achieves an average F1-score of $34.8\% \pm 0.7$.
This result suggests that homophily is the most powerful predictor of Donald Trump support among the considered ones: the role of shared social groups outranks in predictive power direct online interactions, social feedback, bag-of-words, and sentiment-based features.
This result confirms the importance of homophily as a determinant of social behavior~\citep{dellaposta2015liberals}.
We show which specific topical groups are most predictive of Trump support in Section~\ref{sec:feature-analysis}.

\spara{Social feedback.} Reddit scores obtain an F1-score of $33.7\% \pm 1.0$, almost as high as participation.
We remark that, in order to disentangle as much as possible participation and scores, we take the population average score for the subreddits a given user did not participate in.
Therefore, such a high score suggests a relevant role for social feedback and conformity: individuals that were positively or negatively welcomed by certain communities land on \donald four years later.
We look at which community's feedback has this effect in Section~\ref{sec:feature-analysis}.
The independence of scores and participation is confirmed by the increase in F1-score when using both feature sets together, as we show at the end of this section.

While for the other feature sets the best classifier is logistic regression, for score-related features random forest has a better outcome.
Since random forest is a non-linear classifier, its advantage suggests a non-linear relationship between Reddit scores and the likelihood of supporting Donald Trump.

\begin{figure}[t]
  \includegraphics[width=0.6\linewidth]{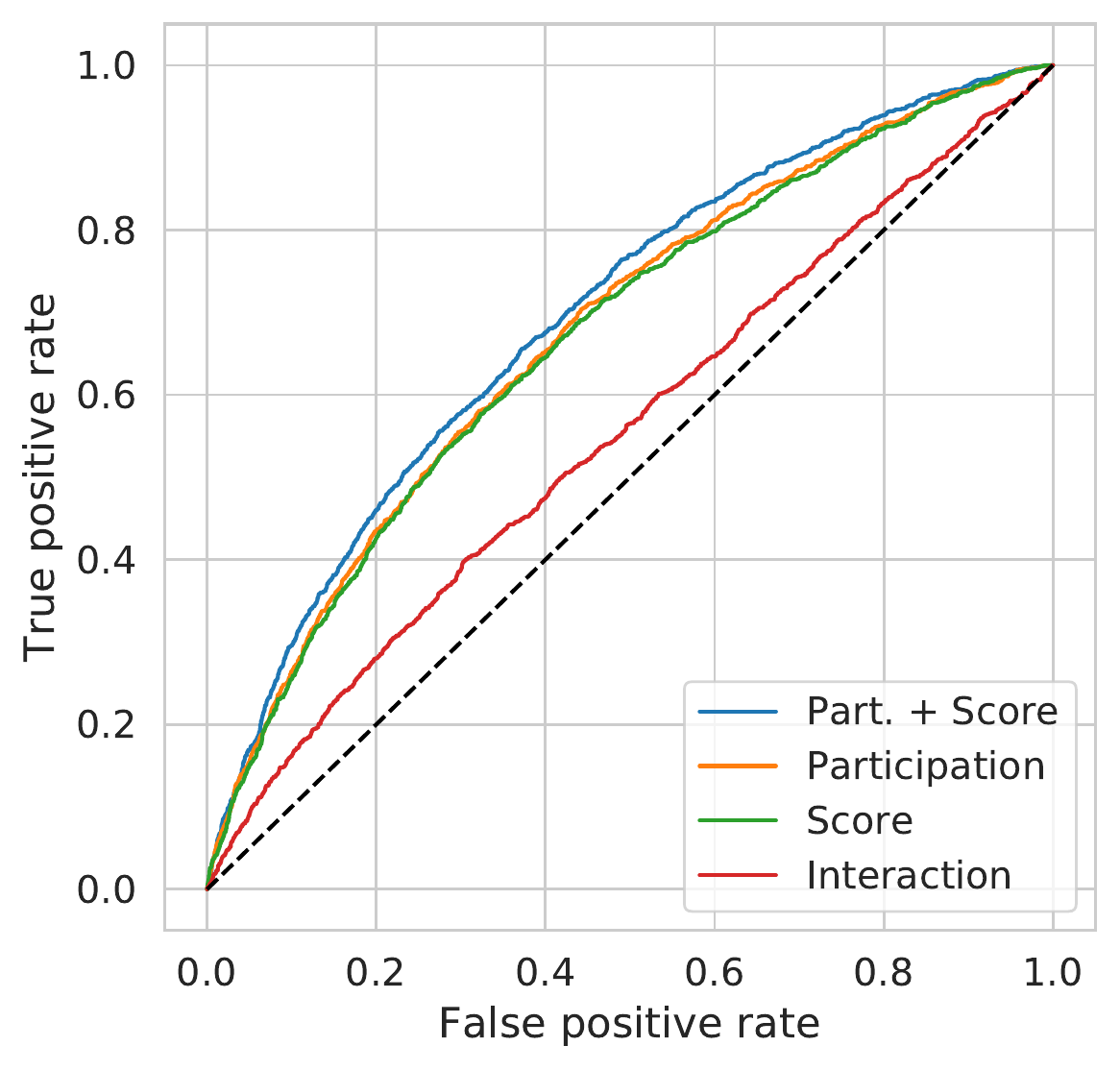}
  \caption{ROC curves of the most predictive feature sets: participation, scores, direct interaction; and the combination of participation and scores. We only report the performance obtained by the best algorithm among logistic regression and random forest. All classifiers use information from 2012 to predict Trump support in 2016.}
  \label{fig:roc-curve}
  \savespace
  \savespaceextra
\end{figure}

\spara{Direct influence.} The effect of interactions, with an F1-score of $26.7\% \pm 0.7$, seems to be much lower than the one of scores and participation.
By using a class-proportional random baseline, we obtain an F1-score of 15.2\% (close to 15.8\%, the proportion of Trump supporters).
Direct interactions are therefore still a better predictor than random.
We investigate in depth the correlations discovered on direct interactions by using the bisected interaction feature set at the end of this section, and by analyzing which are the most important features in Section~\ref{sec:feature-analysis}.

\spara{Language.} Finally, linguistic features perform quite poorly. 
Sentiment, with an F1-score of $16.4\% \pm 13.4$ is as predictive as the random baseline, and any classifier more complex than a decision tree ends up overfitting.
In other words, we do not observe any correlation between the tone of writing and the likelihood of becoming a Trump supporter.
The bag-of-words features perform better, but with $25.9\% \pm 1.0$ of F1-score they are much worse than participation, and still worse than interaction.
This result suggests that simple language models are worse predictors of Trumpism than common social groups.

\begin{table}[t]
  \centering
  \caption{For each of the most predictive feature sets, we report precision, recall, F1-score, and area under ROC curve. We only report the performance obtained by the best algorithm between logistic regression and random forest. All classifiers are 5-fold cross-validated and use information from 2012 to predict Trump support in 2016.}
  \label{table:metrics}
  \savespaceextra
  \begin{tabular}{@{}lrrrr@{}}
    \toprule
                  & Precision & Recall & F1   & AUC  \\
    \midrule
    Participation & 0.25      & 0.56   & 0.34 & 0.68 \\ 
    Score         & 0.24      & 0.60   & 0.33 & 0.67 \\
    Interaction   & 0.18      & 0.52   & 0.26 & 0.55 \\
    Part. + Score & 0.27      & 0.56   & 0.35 & 0.70 \\
    \bottomrule
  \end{tabular}
  \savespace
\end{table}

\mpara{Combined features.} Now, we measure the predictive power of pairs of feature sets used together: participation and scores, participation and interactions, and interactions and scores.
Results show that, first, adding the interaction feature set to any other one does not improve their predicting power.
The results for participation and interactions are the same as those for participation, and for interactions and scores are also the same as those of scores only.
These results strengthen our conclusion that direct online interactions on Reddit are not a decisive factor in determining who becomes a Trump supporter four years later.
Instead, when we combine participation and scores, results improve slightly compared to the best of the two.
This fact suggests that these two types of interactions provide a partially orthogonal signal.
The most important signals we find are therefore homophily and social feedback, while we find only limited effects of social influence.
Combining participation and score thus constitutes our best social features-based classifier.

\smallskip

We analyze in detail the performance of this last model in predicting Trump support four years in advance.
This model obtains a precision of 27\% and a recall of 56\%.
Let us remind that the fraction of Trump supporters in our focus group is 15.8\%.
By taking the probability assigned by the best classifier to each user we obtain a score indicating the propensity of a Reddit user to become a Trump supporter.
We evaluate the predictive power of this propensity score with a ROC curve in Figure~\ref{fig:roc-curve}.
The area under ROC curve for this model is 0.70.
We report these results, along with the models for participation, scores, and interactions taken individually, in Table~\ref{table:metrics}.

\mpara{Bisected features.} We now turn our attention to bisected features.
Recall that by bisecting we mean dividing the subreddits in a certain feature set (Scores or Interactions) in two groups, depending on whether a subreddit has a fraction of future Trump supporters larger ($T$) or smaller ($N$) than average.
As such, these features contain future information, not originally available in 2012, but have a coarser granularity.
They allow us to investigate the effect of influence of (future) Trump-supporting users in contrast with the rest, both for direct influence and social feedback.
First, we measure their results in terms of prediction accuracy, by looking at Table~\ref{tab:main-results}.
Bisected interactions obtain a similar performance to interactions divided by subreddits.
This finding suggests that the effect of social influence is fairly similar across Trump-dominated subreddits.
Surprisingly, instead, scores lose predictive power.
Apparently, the coarser granularity makes the classifier less precise.
This result shows that the effect of social feedback from a certain community is not simply a reflection of whether that community will become more or less dominated by Trump supporters, but there is a finer-grained structure to it.

\medskip

We analyze in depth the features for the two bisected models (scores and direct interactions) in order to further characterize which types of interactions anticipate Donald Trump support.

\smallskip

Let us first look at the logistic regression coefficients for the features in the bisected interaction feature set.
Here we have three features, depending on the interaction being conflictual or non-conflictual, and on it involving a subreddit with a high or low number of future Trump supporters.
Using this kind of future information allow us to look for evidence of backfire effect.
Table~\ref{table:group-interactions} shows that having \emph{any} direct interaction with future Trump-dominated subreddits is predictive of Trump support.
In addition, conflictual interactions (irrespective of the target) are correlated with Trump support, as shown by the negative coefficient for non-conflictual interactions.
This finding is a manifestation of quarreling behavior in Trump supporters online, more than of backfire effect.
This interpretation is consistent with previous analyses~\cite{merrin2019president} and supported by the results we show in the next paragraph.

\begin{table}[t]
  \centering
  \caption{Logistic regression coefficients for predicting Trump support, for all the features in the bisected interaction feature set. We indicate with $T$ the set of subreddits with more Trump supporters than average and with $N$ those with fewer Trump supporters than average.}
  \label{table:group-interactions}
  \begin{tabular}{@{}lr@{}}
    \toprule
    Feature description  & $\beta$  \\
    \midrule
                    {\small Interactions with users participating in $T$} &  0.076163 \\
    {\small Non-conflictual interactions with users participating in $T$} & -0.005322 \\
    {\small Non-conflictual interactions with users participating in $N$} & -0.029029 \\
    \bottomrule
  \end{tabular}
  \savespace
\end{table}

\begin{figure}
  \includegraphics[width=0.85\linewidth]{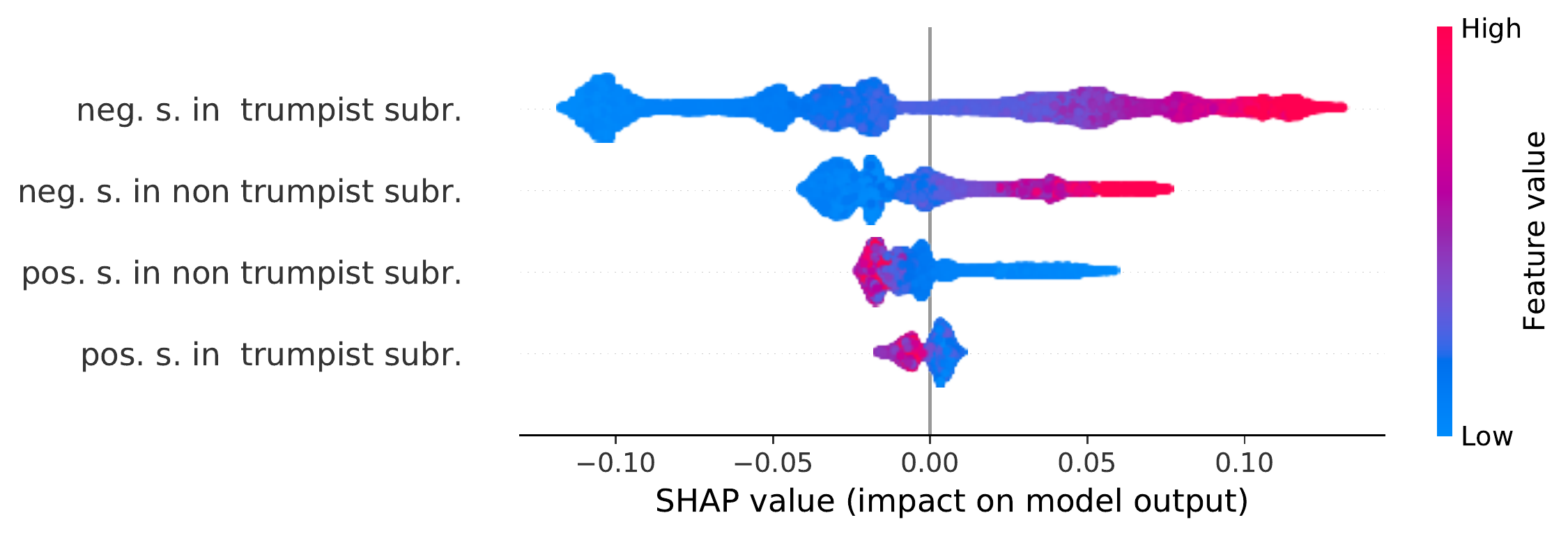}
  \caption{SHAP values for all the features in the bisected scores feature set.
   We indicate with \emph{pos.} the features obtained from positive scores and \emph{neg.} for negative scores. For each feature, red indicates the highest values and blue the lowest. On the right, we have the feature values most associated with Trump support. }
  \label{fig:scores-group}
  \savespace
\end{figure}

\medskip

Second, we analyze the results for the bisected scores feature set.
In this feature set, we divide the subreddits in two groups, according to the number of future Trump supporters.
Therefore, considering positive and negative scores, we have four features.
Since the best classifier for this feature set is random forests we use SHAP, a state-of-the-art algorithm to explain features in random forests models~\cite{shap2, shap1}.
These values can be interpreted similarly to the $\beta$ coefficients of the logistic regression.
Each point represents a user, thus, for each feature, the figure shows the distribution of SHAP values across the data set.
Horizontally wider distributions indicate a larger absolute impact of the feature in the overall classification, while the color of each point (blue to red) encodes the feature value (low or high).
A feature with high values corresponding to positive SHAP values (to the right) is positively correlated with Trumpism.
Conversely a feature with high values corresponding to negative SHAP values (to the left) is negatively correlated Trump support.
We report SHAP values in Figure~\ref{fig:scores-group}.

The results are quite insightful: negative scores in subreddits with higher-than-average future presence of Trump supporters are associated with future Trump support.
It would appear, therefore, that the defiance of social group norms that anticipate Trump support is present also in the communities more aligned with Trumpism.
This is consistent with other findings of ``trolling'' behavior from Trump supporters~\cite{merrin2019president}.

\subsection{Predictive traits}
\label{sec:feature-analysis}

\begin{table}[t]
  \centering
  \caption{
  Logistic regression coefficients for the most important features in the bag-of-words feature set.
  On the left we have the top 10 features with largest $\beta$ coefficient; on the right, the top 10 with smallest $\beta$ coefficient.}
  \label{table:bag}
  {\small
    \begin{tabular}{@{}lrclr@{}}
      \toprule
      \multicolumn{2}{c}{\textbf{Trump supporters}} && \multicolumn{2}{c}{\textbf{Trump non-supporters}} \\
      \cmidrule(lr){1-2} \cmidrule(lr){4-5}
      Word & $\beta$ && Word & $\beta$ \\
      \midrule
      liberal     &  0.000784 && abuse        & -0.000399 \\
      guy         &  0.000691 && reporter     & -0.000345 \\
      debate      &  0.000650 && similar      & -0.000338 \\
      politic     &  0.000635 && contribution & -0.000326 \\
      libertarian &  0.000604 && century      & -0.000322 \\
      come        &  0.000604 && honor        & -0.000321 \\
      think       &  0.000593 && palestinian  & -0.000318 \\
      cop         &  0.000591 && writer       & -0.000314 \\
      tell        &  0.000587 && context      & -0.000313 \\
      home        &  0.000570 && voting       & -0.000306 \\
      \bottomrule
    \end{tabular}
  }
  \savespace
\end{table}

\begin{table}
  \centering
  \caption{Logistic regression coefficients for the most important features in the participation feature set.
  On the left we have the top 30 features with largest $\beta$ coefficient; on the right, the top 30 with smallest $\beta$ coefficient.}
  \label{table:participation}
  \small
  \begin{tabular}{@{}lrlr@{}}
    \toprule
    \multicolumn{2}{c}{\textbf{Trump supporters}} & \multicolumn{2}{c}{\textbf{Trump non-supporters}} \\
    \cmidrule(lr){1-2} \cmidrule(lr){3-4}
    Subreddit & $\beta$ & Subreddit & $\beta$ \\
    \midrule
    \R{Conservative}         &  0.3815 & \R{raspberry\_pi}     & -0.2847 \\
    \R{Libertarian}          &  0.3740 & \R{TrueAtheism}       & -0.2577 \\
    \R{conspiracy}           &  0.3733 & \R{AskCulinary}       & -0.2355 \\
    \R{4chan}                &  0.3341 & \R{comics}            & -0.2249 \\
    \R{circlejerk}           &  0.3107 & \R{rpg}               & -0.2186 \\
    \R{NoFap}                &  0.2918 & \R{ireland}           & -0.2034 \\
    \R{Entrepreneur}         &  0.2539 & \R{Fantasy}           & -0.1983 \\
    \R{ImGoingToHellForThis} &  0.2510 & \R{explainlikeimfive} & -0.1944 \\
    \R{trees}                &  0.2482 & \R{environment}       & -0.1892 \\
    \R{MensRights}           &  0.2482 & \R{doctorwho}         & -0.1878 \\
    \R{guns}                 &  0.2293 & \R{polyamory}         & -0.1806 \\
    \R{blackops2}            &  0.2110 & \R{scifi}             & -0.1777 \\
    \R{runescape}            &  0.2031 & \R{books}             & -0.1772 \\
    \R{Anarcho\_Capitalism}  &  0.1937 & \R{askscience}        & -0.1738 \\
    \R{Catholicism}          &  0.1931 & \R{london}            & -0.1691 \\
    \R{leagueoflegends}      &  0.1920 & \R{britishproblems}   & -0.1687 \\
    \R{nfl}                  &  0.1843 & \R{Homebrewing}       & -0.1632 \\
    \R{starcraft}            &  0.1714 & \R{programming}       & -0.1521 \\
    \R{CCW}                  &  0.1638 & \R{gadgets}           & -0.1501 \\
    \R{breakingbad}          &  0.1631 & \R{AndroidQuestions}  & -0.1463 \\
    \R{investing}            &  0.1624 & \R{listentothis}      & -0.1462 \\
    \R{AdviceAnimals}        &  0.1589 & \R{hiphopheads}       & -0.1397 \\
    \R{DeadBedrooms}         &  0.1577 & \R{boardgames}        & -0.1336 \\
    \R{Firearms}             &  0.1551 & \R{asoiaf}            & -0.1292 \\
    \R{Advice}               &  0.1537 & \R{whatisthisthing}   & -0.1244 \\
    \R{seduction}            &  0.1518 & \R{lgbt}              & -0.1187 \\
    \R{Christianity}         &  0.1455 & \R{cringepics}        & -0.1175 \\
    \R{golf}                 &  0.1453 & \R{ukpolitics}        & -0.1136 \\
    \R{mylittlepony}         &  0.1437 & \R{Python}            & -0.1089 \\
    \R{POLITIC}              &  0.1423 & \R{baseball}          & -0.1080 \\
    \bottomrule
  \end{tabular}
  \vspace{-4mm}
\end{table}

In this section, we investigate the importance of each feature for our models, in order to answer our last research question: which traits did anticipate the development of Donald Trump support?
To do so, we perform an in-depth feature analysis for the most successful models: bag-of-words, participation, scores, interactions, and the combined model.

As seen in the last section, the best classification algorithm is in general the logistic regression; for the scores feature sets, random forests achieve similar or better performance, possibly because of their non-linearity.
Therefore, in our investigation of feature importance, we analyze random forests features when scores are involved, and logistic regression otherwise.
Thanks to the normalization described in Section~\ref{sec:methods}, for logistic regression we can simply look at the coefficients obtained by each feature.
Instead, for random forests, we employ again SHAP, an algorithm to explain the output of ensemble tree models~\cite{shap2, shap1}.

\spara{Language features.} The first model we investigate is the bag-of-words model.
The model tries to capture statistical differences in the usage of words by Trump supporters.
Table~\ref{table:bag} reports the most discriminative words.
In general, these features are not easily interpretable, but we can discern some noticeable patterns.

Trump supporters in 2012 were more likely to use the word \emph{liberal} and the word \emph{libertarian}.
We can surmise that the former is an insult and the second is a self-description, but there is no direct way to confirm this conjecture by looking at the model alone.
However, we shall see some confirmatory evidence in the analysis of participation features.
Moreover, they use terms such as \emph{cop}, possibly linked to the law-and-order views promoted by Trump; and \emph{home}, perhaps related to a pronounced attention to concepts such as family values, or homeland.

On the opposite side --the words least used by Trump supporters in 2012-- we note terms vaguely related to civil rights such as \emph{abuse}, \emph{reporter}; and the word \emph{palestinian}, possibly acknowledging claims of Palestinians.
However, in general also the features on this side are hard to interpret.
We shall now see how, by using the more predictive participation-based classifier, we are able to draw a clearer portrait.


\spara{Participation features.}
We have seen that this is the best single feature set in terms of prediction accuracy.
Table~\ref{table:participation} shows the 30 most important features for each of the two classes.
Here, each feature represents participation (writing a comment) in that subreddit in 2012.
The model coefficients are larger than for the bag-of-words features.

The most discriminative features are related to political views.
Conservative and libertarian groups are the most correlated with Donald Trump support.
This finding is consistent with the idea that Trump's coalition is a part of the so-called ``libertarian authoritarianism'', which conflates needs from both ideological camps~\cite{brown2018fires}.

We also recognize topics and communities that are known to be associated with Trump support.
\R{conspiracy} is a community devoted to conspiracy theories~\cite{klein2019pathways}; e.g., it covered extensively the ``pizzagate'' hoax about child sex rings operated by Democratic party officials.
This observation backs the theory that some fringe groups have merged into the mainstream political discourse~\cite{nithyanand2017online}.

The website 4chan, a ``politically incorrect'' discussion board, has been linked to the ``alt-right movement'' in a previous analysis~\cite{merrin2019president}.
We find that participation to the \R{4chan} subreddit in 2012 is the fourth most predictive feature in this set.
Other politically incorrect groups are also correlated with Trump support.
For example, \R{ImGoingToHellForThis} is a community devoted to shocking and vitriolic humor.

Some interests and hobbies clearly emerge among the most predictive subreddits for Trump support, while others seems to anti-correlate with Trump support.
An interest in firearms is strongly correlated with Trumpism (\R{guns}, \R{Firearms}, \R{CCW} [Concealed Carry Weapons]).
The same is true for several video games communities (\R{blackops2}, \R{runescape}, \R{leagueoflegends}, \R{starcraft}).
Instead, other hobbies are anti-correlated, for instance, tabletop games (\R{boardgames}, \R{rpg}).
Cuisine and do-it-yourself hobbies are among the most important: \R{raspberry\_pi}, \R{AskCulinary}, \linebreak[4] \R{Homebrewing} are strongly anti-correlated with Trump support.
Interests in literature and art is an equally important predictor (\R{books}, \R{comics}, \R{ListenToThis}, \R{Fantasy}, \R{scifi}).

Religion is also central in the separation: among those correlated with Trump support we find \R{Catholicism} and \R{Christianity}; among those anti-correlated, instead, one of the most predictive is \R{TrueAtheism}.
This finding is consistent with the idea that, for many Americans, Trump was ``a symbolic defense of the United States perceived Christian heritage''~\cite{whitehead2018make}.

Some of the communities correlated with Trump support are related to interests such as entrepreneurship and investing. This could suggest both support from wealthy persons, or from those with a self-made attitude.
Status threat (as opposed to economic hardship) has been indicated as a common trait in Trump support~\cite{mutz2018status}.

\begin{figure}[t]
  \includegraphics[width=0.75\linewidth]{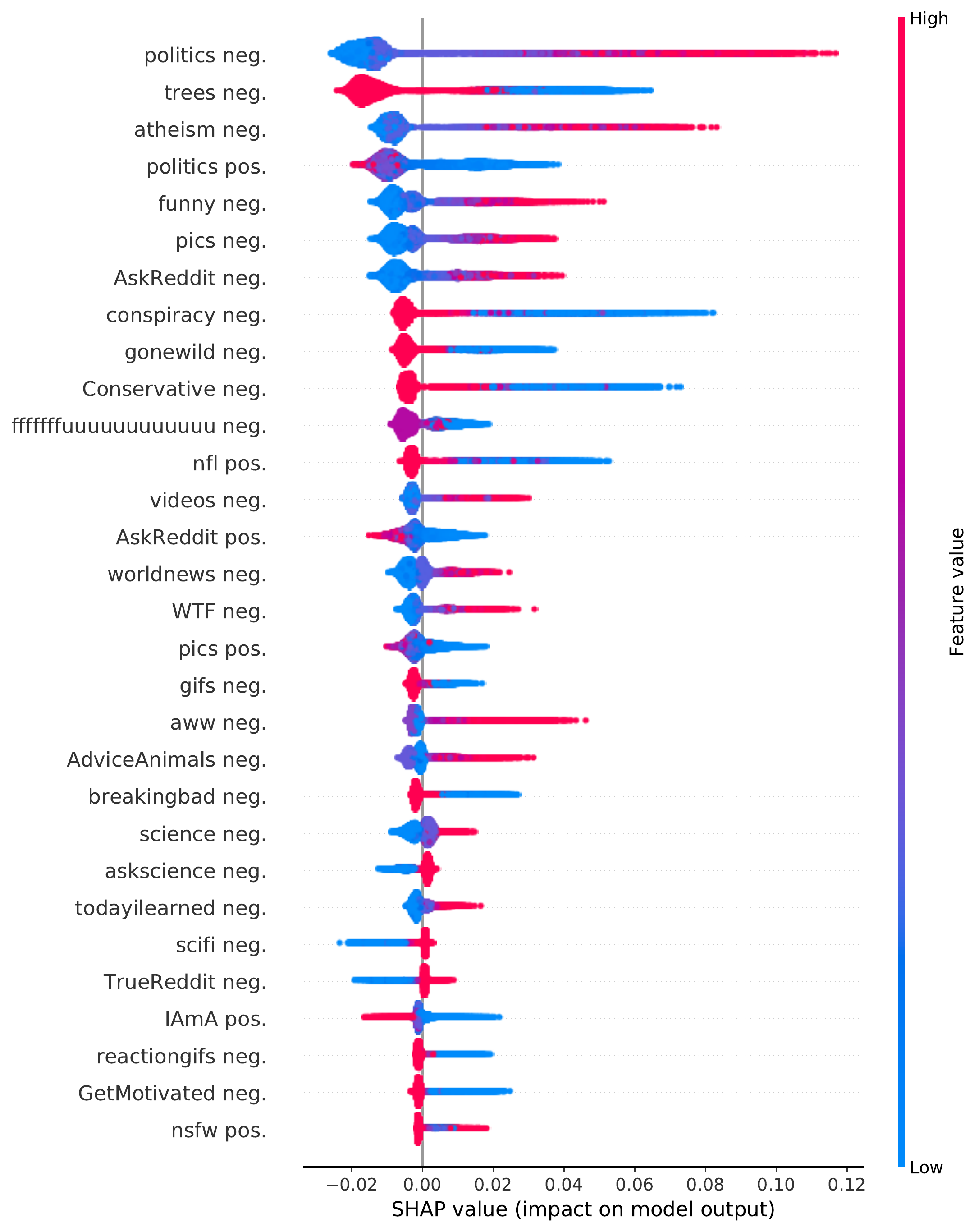}
  \caption{SHAP values for the 30 most important features in the score feature set.
   We indicate with \emph{pos.} the features obtained from positive scores and \emph{neg.} for negative scores. For each feature, red indicates the highest values and blue the lowest. On the right, we have the feature values most associated with Trump support. For instance, the first row indicates that a high negative score in \R{politics} is indicative of Trump support.}
  \label{fig:scores}
  \savespace
\end{figure}

Several subreddits with predominantly male demographics appear among those correlated with Trump support, consistently with previous findings~\cite{brewer2019ground}.
One of them, \R{MensRights}, is focused on the defense of male interests against feminism.
From a sexual orientation point of view, we observe a very clear division between Trump-associated subreddits and the anti-correlated ones.
The latter group includes gender, sexual, and romantic minorities, such as \R{polyamory} and \R{lgbt}. 
The subreddits most positively correlated with Trump are mostly masculine: for instance, \R{seduction}, a subreddit part of the Pick-Up Artists movement;\footnote{https://www.dailydot.com/irl/ken-hoinsky-pua-reddit-seduction-book-the-game} \R{NoFap}, a group that provides self-help for porn addiction; and the already cited \R{MensRights}.
It is worth noting that also \R{DeadBedrooms}, which self-describes as ``a support group for Redditors who are coping with a relationship that is seriously lacking in sexual intimacy'', is among the most associated with Trump support.

Of the remaining subreddits in the group, many are associated with popular culture (on both sides), such as sports and TV shows. 
Other subreddits appear to be anti-correlated with Trump support simply because they are typically associated to non-American Reddit users: this is the case for \R{ukpolitics}, \R{london}, \R{ireland}, \R{britishproblems}.
A curious finding is that one of the best predictors for Trump support is \R{trees}, a subreddit for cannabis enthusiasts.  We suspect a possible confounding factor: for instance, \citet{miech2019monitoring} show that, in the United States, daily cannabis usage in 19-24 years olds is three times higher for those who are not attending college (13\% vs 4\%). This is consistent with previous finding that Trump has attracted more support from this less-educated segment of the population~\cite{rothwell2016explaining}.

\mpara{Social feedback.}
We now turn our attention to the social feedback features.
As mentioned before, since the best model for this feature set is random forest, we employ SHAP~\cite{shap2, shap1} to explain the relationships learned by the model.
Figure~\ref{fig:scores} reports the resulting SHAP values. 

Some of the subreddits to which participation is a strong predictor of Trump support also appear here, although in a different guise: negative scores in \R{Conservative}, \R{trees}, and \R{conspiracy} are correlated with lack of support for Trump, while negative scores in \R{atheism} are correlated with Trump support.

On the \R{NFL} subreddit, we observe an anti-correlation between positive scores and Trump support.
Since this subreddit also appears among the most important participation features, this result suggests that participating in the subreddit but not being appreciated by the community is a predictor of Trump support.

Some generalist subreddits, such as \R{funny}, \R{pics}, or \linebreak[4] \R{AskReddit} also appear.
In all these cases, negative scores are associated with Trump support; the same is true for \R{politics}.
We remind that participation in those subreddits is not among the most important features.
These observations suggest that a negative feedback from wide-ranging, mainstream Reddit communities in 2012 is linked to Trump support in 2016.

This could be the case also for \R{gonewild}, a subreddit which self-describes as a ``a place for open-minded adult Redditors to show off their nude bodies for fun'': users who obtain negative feedback in this community are more likely to become Trump supporters four years later.

\begin{table}
  \footnotesize
  \centering
  \caption{Logistic regression coefficients for the most important features in the interaction feature set.
  On the left we have the top 10 features with largest $\beta$ coefficient; on the right, the top 10 with smallest $\beta$ coefficient.
  We indicate with ``$r$ dist.'' the fraction of interactions on subreddit $r$, and with ``$r$ pos.'' the fraction of positive interactions over all interactions with subreddit $r$.}
  \label{table:interaction}
  \begin{tabular}{@{}lrlr@{}}
    \toprule
    \multicolumn{2}{c}{\textbf{Trump supporters}} & \multicolumn{2}{c}{\textbf{Trump non-supporters}} \\
    \cmidrule(lr){1-2} \cmidrule(lr){3-4}
    Feature & $\beta$ & Feature & $\beta$ \\
    \midrule
    \Rsmall{ShitPoliticsSays} pos. &  0.1513 & \Rsmall{todayilearned} dist.       & -0.0698 \\
    \Rsmall{Republican} pos.       &  0.0868 & \Rsmall{TrueReddit} dist.          & -0.0584 \\
    \Rsmall{conspiracy} dist.      &  0.0684 & \Rsmall{Futurology} dist.          & -0.0548 \\
    \Rsmall{moderatepolitics} pos. &  0.0637 & \Rsmall{dataisbeautiful} dist.     & -0.0348 \\
    \Rsmall{Conservative} dist.    &  0.0563 & \Rsmall{GaryJohnson} dist.         & -0.0214 \\
    \Rsmall{Libertarian} dist.     &  0.0543 & \Rsmall{PoliticalDiscussion} dist. & -0.0168 \\
    \Rsmall{Libertarian} pos.      &  0.0457 & \Rsmall{Liberal} dist.             & -0.0149 \\
    \Rsmall{conspiracy} pos.       &  0.0416 & \Rsmall{PoliticalDiscussion} pos.  & -0.0120 \\
    \Rsmall{POLITIC} pos.          &  0.0274 & \Rsmall{worldnews} dist.           & -0.0116 \\
    \Rsmall{Economics} pos.        &  0.0219 & Total number of interactions       & -0.0105 \\
    \bottomrule
  \end{tabular}
  \savespace
\end{table}

\spara{Direct influence.}
Our third basic feature set represents direct interactions between a user and another user, where the latter participated in a certain subreddit.
They also account for how many of those interactions were non-conflictual.
Table~\ref{table:interaction} shows the most predictive features.
Despite its scarce predictive power when compared to participation, we are still able to use these features to enrich our portrait.

Trump support is predicted by the fraction of positive interactions on politically-active subreddits such as \R{Republican}, \linebreak[4] \R{Libertarian}, and \R{moderatepolitics}, as well as communities which discuss topics of interests to Trump supporters such as \R{conspiracy} and \R{Economics}.
These traits support our previous analysis, and confirm the idea that libertarianism and conservatism are among the roots of Trumpism.
However, we also observe that the amount of interactions with \R{GaryJohnson}, candidate against Trump in 2016 elections, is anti-correlated with Trump support.
The most powerful feature in this set is the fraction of positive interactions on \R{ShitPoliticsSays}.
This subreddit hosts critiques and mockery of other subreddits, and it exhibits right-wing views.\footnote{E.g., it denounces \R{Fuckthealtright} and \R{AgainstHateSubreddits} as hostile subreddits.}

Finally, we note that the total number of direct interactions is anti-correlated with Trump support, suggesting that the overall influence of Reddit is adverse to Trump.

\begin{figure}
  \includegraphics[width=0.75\linewidth]{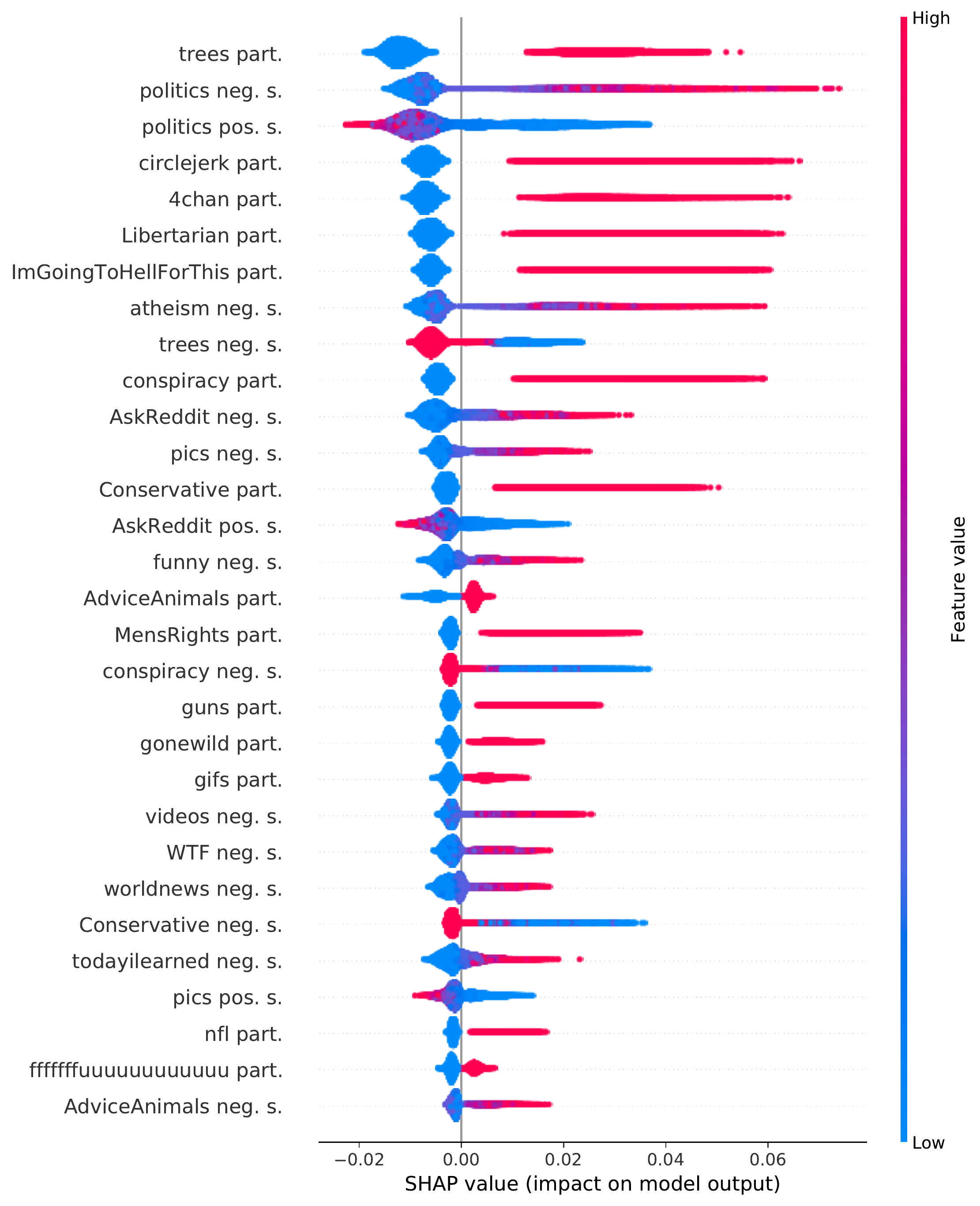}
  \caption{SHAP values for the 30 most important features in the participation+score feature sets combined.
   We indicate with \emph{pos. s.} the features obtained from positive scores and \emph{neg. s.} for negative scores; with \emph{part.} the participation features. For each feature, red indicates the highest values and blue the lowest. On the right, we have the feature values most associated with Trump support.}
  \label{fig:part-score-int}
  \savespace
  \savespaceextra
\end{figure}

\spara{Combined features.}
Finally, Figure~\ref{fig:part-score-int} displays the most important features of the combined model that uses participation and scores.
The two feature sets are well balanced: both feature sets are represented among the most predictive features (14-to-16).
This observation strengthens the hypothesis that social feedback and homophily provide a different, orthogonal signals in predicting support for Trump.

\section{Conclusions and Future Work}
\label{sec:conclusions}

We have looked at predictors for becoming a supporter of Donald Trump on Reddit.
We used data from 2012 to predict the participation in \donald in 2016, which we use as a proxy for support of Trump.
Such a prediction task is challenging, given the four-year time span (a US presidential electoral cycle) between the observed data and the target behavior.
Nevertheless, our best performing model achieves an AUC of $0.70$ and an F1 measure of $0.36$, significantly above the performance of a random baseline.

We explored a diverse set of predictors which represent three sociological hypotheses for the support of Trump: homophily, social feedback, and influence.
We operationalized each hypothesis in the context of Reddit by looking at participation of a user in a community (a subreddit), the appreciation their posts receive in a given community, and interactions with users of other communities.
Compared to other baseline interpretable linguistic features, such as the bag-of-words and the sentiment of the posts, the social ones result more predictive of the target behavior.
In particular, features encoding homophily and social feedback (conformity and anti-conformity) have shown to be the best predictors of Trumpism, while social influence has shown limited relevance.
In addition, a combination of features for homophily and social feedback (i.e., participation and scores) performs slightly better than the single features, thus showing that the two signals are somewhat complementary.

Finally, we introspect the features of the best performing models to delineate a `persona' of how a typical Trump supporter in 2016 looked like on Reddit in 2012.
The typical Trump supporter has conservative and libertarian views, is ill-received by the mainstream political tribe, is religious and in conflict with atheism, and has interests in guns, conspiracies, entrepreneurship, and politically incorrect content.
Conversely, the typical Reddit user who does not support Trump is atheist, LGBT-friendly, and has interests in cooking, literature, and technology.

\paragraph{Limitations and future work.}

The operationalization of the sociological theories we considered in this study has, necessarily, the opportunity to introduce distortions. 
Out of the three feature sets, the interaction ones which encode social influence are the most brittle because of their natural sparsity.
We countered this characteristic by aggregating them per community, but they still resulted to be the least predictive ones in our models.
This result might be caused by the specific design choices, and more work is needed to quantify the role that social influence plays in changing the political attitudes of people on social media.

The score feature set which encode social feedback also presents some challenges, as the score distribution is heavy tailed.
In our work, we used a non-linear classifier (random forest) to tackle this problem, but more sophisticated algorithms might improve results.

More fundamentally, the design of the current study does not allow to differentiate between different causal interpretations of the social feedback effect.
Let us use three variables to represent the behavior of supporters: observed social feedback, observed support for Trump, and latent political attitudes.
On the one hand, a causal model could envision the social feedback as a cause for change in political attitudes, which in turn causes the support for Trump.
In this case, the social feedback is a root \emph{cause} of the support for Trump.
For instance, a user might have a negative experience with the mainstream political community, which causes their attitudes to drift towards more extreme positions, which in turn might explain the support for Trump.
On the other hand, the latent political attitudes could be a common cause for both the received social feedback, because the attitudes expressed are already misaligned with the community, and the support for Trump.
In this second case, the social feedback is an \emph{effect} of the political attitudes, and the support for Trump depends on it in a non-causal way.
For example, a user might have some fringe attitudes which are ill-received in the mainstream political community, and find a natural outlet in Trumpism.
A causal investigation of these hypotheses from observational data is an interesting extension of the current work~\citep{pearl2009causality}.
In this framework, we could formalize confounding factors, understanding for instance if Trump supporters became more engaged with some political subreddits, or if they stem from users more active on them in the first place.
However, our work constitutes a necessary first step before any causal investigation.

%

Finally, we have described the `persona' of a Trump supporter by assuming there is only a single one.
However, there is evidence that people coming from multiple socio-demographics strata support Trump~\citep{manza2017working}.\footnote{\url{https://fivethirtyeight.com/features/the-mythology-of-trumps-working-class-support}}
It is thus possible that the persona we describe is an amalgamation of traits coming from different sources.
In this case, building multiple personae would create more accurate portraits.
Also, it would help in distinguishing Trump supporters on Reddit from other young U.S. Republicans.
This analysis could help understand which issues attracted those who became politicized in this way, thus giving more insights on the roots of Trumpism.

\vspace{7mm} 

\balance
\bibliographystyle{ACM-Reference-Format}
\bibliography{references}

\end{document}